# Microstructure and mechanical properties of Al-Cu alloy with 0.6%Fe produced with ultrasonic vibration and applied pressure

Yuliang Zhao, Weiwen Zhang*, Fansheng Meng, Datong Zhang, Chao Yang, Zhiyu Xiao

(School of Mechanical and Automotive Engineering, South China University of Technology, Guangzhou, 510640, China)

*Corresponding author. Tel:+86-20-87112272, Fax:+86-20-87112111.

E-mail: mewzhang@scut.edu.cn

**Abstract:** The combined effect of ultrasonic vibration (UT) and applied pressure (P) on microstructure and mechanical properties of Al-5.0Cu-0.6Mn-0.6Fe alloy were investigated. The best tensile properties produced by P+UT processing are UTS: 268MPa, YS: 192MPa, E.L.: 17.1%, respectively, which increasing by 64%, 59% and 307%, respectively, compare to the Non-treated alloy.

1. Introduction

The Al-Cu based cast alloys have been widely used in the transportation, aerospace and military industry owing to their excellent mechanical properties and low cost. However, iron is common and inevitable impurity element in Al-Cu alloys, especially in the recycled Al-Cu alloys. Owing to the low solubility in Al-Cu alloys[1], Fe usually exists in the Al-Cu alloys in the form of Fe-rich intermetallics which are very brittle and lead to a deleterious effect on the mechanical properties of Al-Cu alloys[2]. It is well known that the morphology, size and distribution of Fe-rich intermetallics have a great influence on the mechanical properties. For example, the formation of α-Fe with Chinese script morphology rather than β-Fe with needle like is less harmful to mechanical properties[3]. Therefore, it is important to find the appropriate way to modify the morphology of Fe-rich intermetallics to minimize the detrimental effect.

The application of compound field, i.e., ultrasonic and electromagnetic field, in solidification has attracted great attention recently. Ultrasonic vibration has been proved to be an environmental and effective process to refine the microstructure and improve the mechanical properties during solidification [4]. Haghayeghi et al. [5] and Zhang et al. [6] demonstrated that it is possible to obtain refined and homogeneously microstructure by applying electromagnetic and ultrasonic fields to Al alloy during solidification. By applying indirect ultrasonic vibration into semisolid slurry Al alloy and then direct squeeze casting, Lü et al. [7] found the alloy have uniform microstructure and enhanced mechanical property. The combined magnetic fields can also be used to refine Al alloy, as demonstrated by Haghayeghi et al. [8], who obtained more refined microstructure compare to single field. Squeeze casting is a technology with short route, high efficiency and precise forming, possessing features of casting and plastic processing, which is often used to prepare high performance Al alloys [9]. However, there are few literature studying the effect of ultrasonic vibration on aluminum alloys during squeeze casting. In this article, the effect of combined ultrasonic vibration and applied pressure on the refine the microstructure of Al-5.0Cu-0.6Mn-0.6Fe alloy were investigated, and the resulting mechanical properties evaluated.

## 2. Experimental procedures

The chemical compositions of Al-5.0Cu-0.6Mn-0.6Fe alloys are Cu 5.4%, Mn 0.63%, Fe 0.63% and Al balance (mass%). 6 Kg raw materials were melted at 750 °C in electric resistance furnace by and nitrogen was introduced to minimize hydrogen content. The experimental equipment is illustrated in Fig. 1a. The ultrasonic vibration system consists of a 1000 W generator, 20 kHz transducer and Ti alloys horn. The die which was placed inside the squeeze casting equipment and the temperature of die was set at approximately 200 °C, and the pouring temperature was set at 710 °C. The ultrasonic power is 900 W before the horn was preheated to 600°C and the applied pressure is 50 MPa, it worked simultaneously for 30 seconds. Finally, the samples with the size of 75mm × 75mm × 70mm were obtained.

In order to compare the microstructures and mechanical properties of the alloy at different condition, the samples were cut at different position of the ingots, as shown in Fig. 1b. Samples for metallographic observation were taken from ingot near the horn with the size of Φ10mm×2mm, and they were etched with 0.5ml HF solution for 30 seconds. The diameter of the Fe-rich intermetallics, Second Dendrite Arm Space (SDAS) and volume fraction of porosity were measured by means of a Leica light optical microscopy equipped with the image analyzer. In quantitative stereology, the measured area fraction is assumption equal to the volume fraction. The diameter of Fe-rich intermetallics in this study is defined as:

$$d = 2\sqrt{\frac{area}{\pi}}$$

Where $d$ is the diameter of Fe-rich intermetallics. The tensile test was carried out on a SANS CMT5105 standard testing machine with a strain rate of 1 mm/min. The dimension of the tensile sample in reference [10] was applied in this study.

3. Results and discussion

Fig. 2 shows the OM microstructure of alloys by different methods, such as non-treatment (Non-treated), pressure (P), ultrasonic treatment (UT) and combined ultrasonic treatment and applied pressures (P+UT). It can 4be observed that the coarse Fe-rich intermetallics and high porosities disperse among the fully developed α-Al dendrites in Non-treated alloy (Fig.2a). Fig.2b reveals that the porosity is hardly to see and the α-Al grains are slightly refined under P. A bimodal structure [11] which exhibits distinct regions of fine and coarse dendritic structures is found in the alloy.

Fig.2c presents that the α-Al dendrite becomes more globular and porosity decreases to some extent, which are agreement with the foundation by Puga et al. [12] and Lü et al. [13] at which UT was applied into the melt during solidification. As shown in Fig.2d, the alloy produced by P+UT has a refined microstructure with equiaxed grains and porosity-free, which indicated that this promote the refinement of microstructure and reduction of porosity.

In order to explicit the shape and morphology of Fe-rich intermetallics and α-Al grains, the SEM images of the samples prepared by different methods are shown in Fig.3. The microstructure of the alloy consist of primary α-Al, eutectic phase $Al_2Cu$ and Fe-rich intermetallics which are mainly $Al_{15}(FeMn)_3Cu_2$ (α-Fe) phase and a few $Al_6(FeMn)$ phase also exist. These results found to be similar to those obtain by Lin et al. [10]. Fig.3 also suggests that morphology and size of Fe-rich intermetallics and $Al_2Cu$ can be significantly changed under P+UT, such as, the morphology of α-Fe phase change from Chinese script (Not-treated) to skeleton script (P+UT) and the size of $Al_6(FeMn)$ and $Al_2Cu$ phase with P+UT processing are much smaller than those with Not-treated.

To quantify the diameter of the Fe-rich intermetallics, SDAS and volume percent of porosity in samples produced by different methods, and the results are presented in Fig.4. It is clear from Fig.4a that the diameter of Fe-rich intermetallics of the alloy under P+UT is much lower than those by individual process, i.e. P or UT. As the processing is changed from the Non-treated to P+UT, the SDAS is decrease from 81μm to 30μm, which is around 170% lower than that of Non-treated alloy. Fig.4b shows that the alloy produced by Non-treated, P, UT and P+UT processing has the porosity levels of 2.1%, 0.1%, 0.6% and 0%, respectively, while the alloy produced by combined P+UT has a free of porosity.

The resulting mechanical properties for four conditions are presented in Table 1.

The optimal properties (UTS: 268MPa, YS: 192MPa, E.L.:17.1 %) of the alloy can be achieved by the combined P+UT processing. Compared to the Not-treated alloy, the processing of P+UT play a crucial role on improvement of tensile properties, which are increased by 64%, 59% and 307%, respectively. Furthermore, the mechanical properties of P+UT are also superior to those produced by individual process, i.e., P or UT. This is direct corresponding to the co-work of ultrasonic vibration and applied pressure, leading to the refinement of microstructure, reduction of porosity and modification of Fe-rich intermetallics. It is well known that the mechanical properties of Al-Cu alloys depend on the morphology and size of microstructure [4], as well as, the distribution of porosity [10]. This mechanism can be explained as the following three reasons. Firstly, the large numbers of nucleation sites are generated by high temperature and pressure under the ultrasonic vibration, causes the improvement of nucleation destiny leading to the formation of the globular and refined α-Al grains and modified α-Fe and $Al_6(FeMn)$ phase [4,14]. These play an important role on the improvement of both strength and ductility. Secondly, the combination of ultrasonic vibration with applied pressure provides a higher cooling rate and heat transfer resulting in a refined microstructure [4]. Thirdly, the P+UT processing is favorable to eliminate the defects of solidification shrinkage and porosity [15-16], which have a great impact on mechanical behavior.

  The SEM tensile fractures of different methods are shown in Fig.5. In Fig.5a, the Chinese-script α-Fe formed around the α-Al dendrite which formed the shrinkage porosity area in the Non-treated alloy can be seen clearly. Fracture surface of Fig.5b shows extensive irregular cleavage and tearing ridges which indicates quasi-cleavage fracture under applied pressure. As shown in Fig.5c, it is noted that UT increases the number of dimples, and the fracture surface of this alloy display a mixed of porosity and dimple morphology. Therefore, an obvious morphology of dimple fracture which deep dimples are distributed uniformly in P+UT alloy (Fig.5d), revealing the highest elongation in this alloy.

**4. Conclusion**

1. The combined P+UT processing promote the refinement of α-Al and Fe-rich intermetalics and reduction of porosity of the Al-5.0Cu-0.6Mn-0.6Fe alloy during solidification.

2. The processing of P+UT changes the morphology of α-Al from dendritic to globular shape and Fe-rich intermetalics from Chinese script to skeleton shape.

3. As compare with the Non-treated alloy, the UTS, YS and E.L. of the alloy produced by P+UT processing are increasing by 64%, 59% and 307%, respectively.

**Acknowledgement**

The financial support from the Guangdong Natural Science Foundation for Research Team (015A030312003), Natural Science Foundation of China (51374110) is acknowledged.

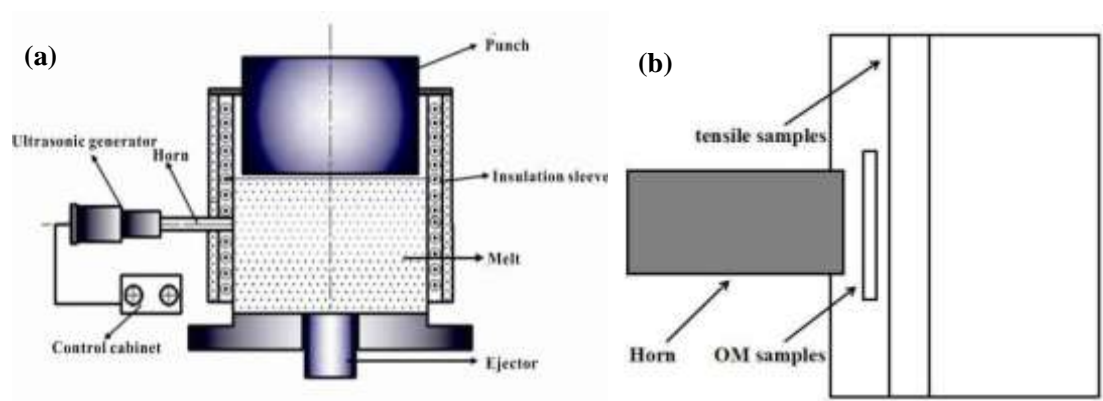

**Fig.1** Schematics view of (a) the combined ultrasonic vibration and squeeze casting equipment; (b) the positions of sample taken from an ingot.

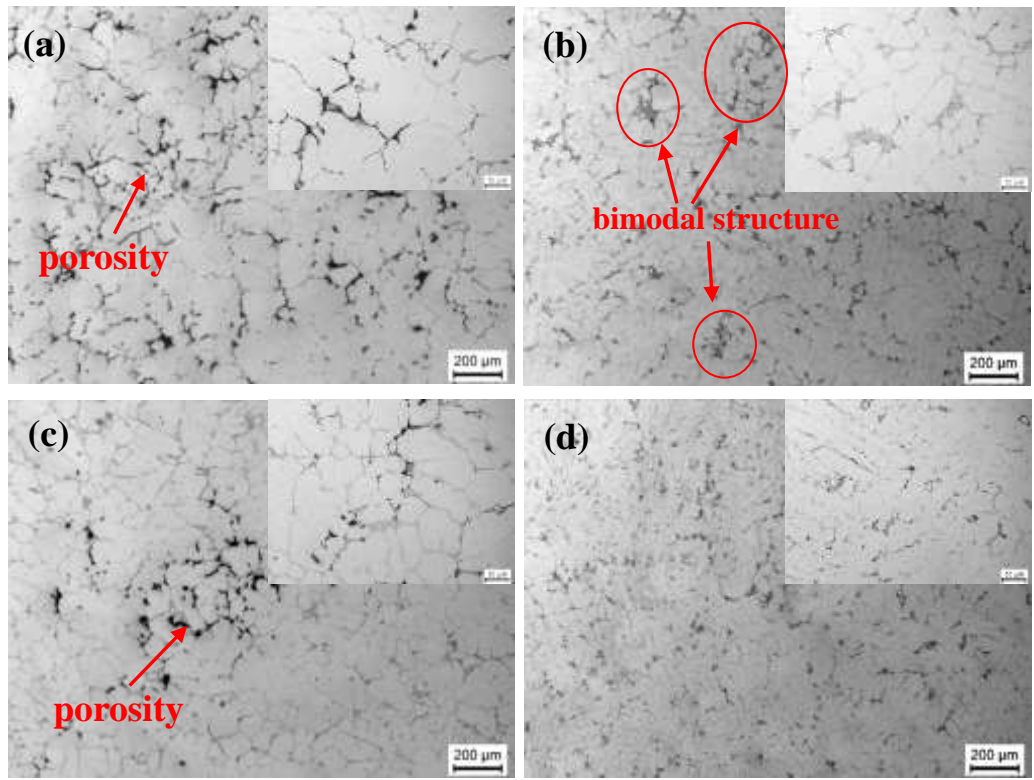

**Fig.2.** Microstructures of Al-5.0Cu-0.6Mn-0.6Fe alloy prepared by different methods: (a) Non-treated; (b) P; (c) UT; (d) P+UT.

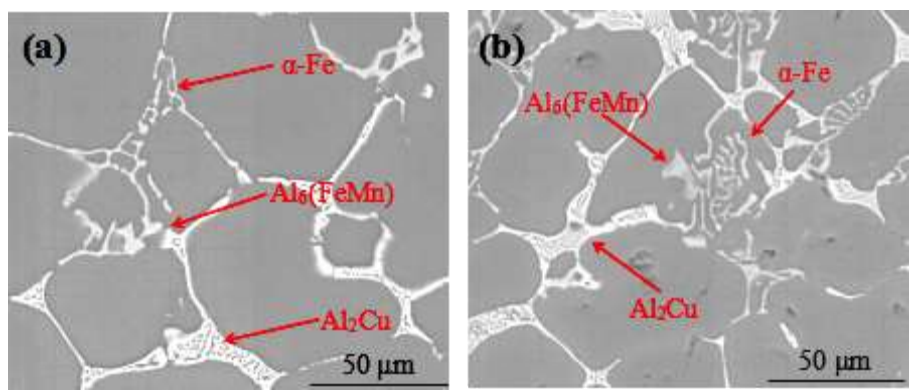

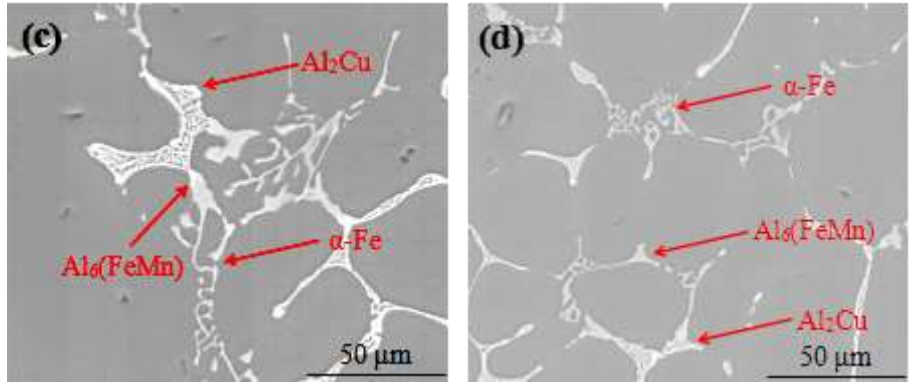

**Fig. 3** SEM images of Al-5.0Cu-0.6Mn-0.6Fe alloys prepared by different methods: (a) Non-treated; (b) P; (c) UT; (d) P+UT.

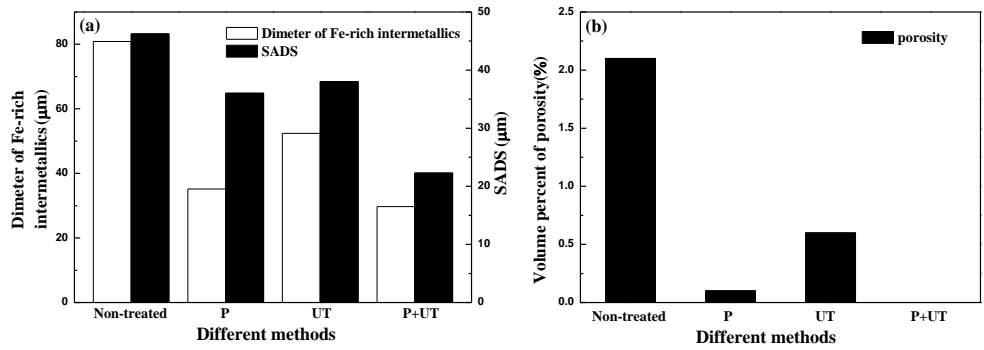

**Fig.4.** Microstructural characteristics of alloys prepared by different methods: (a) diameter of Fe-rich intermetallics and SDAS; (b) Volume percent of porosity.

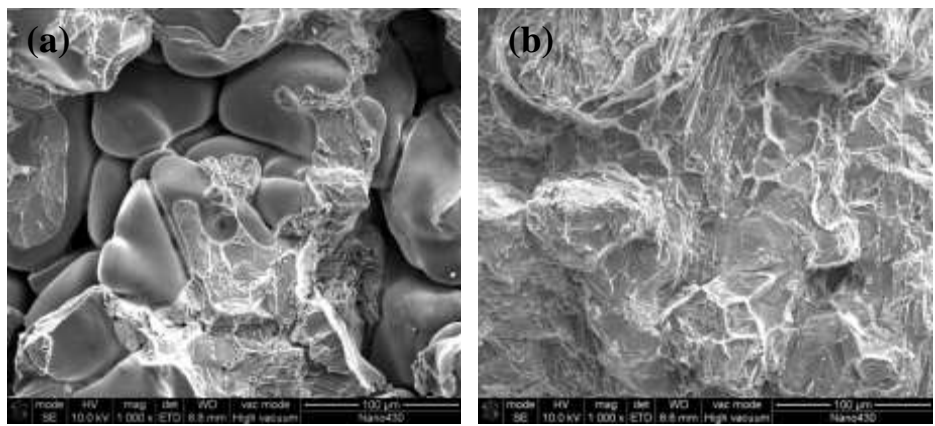

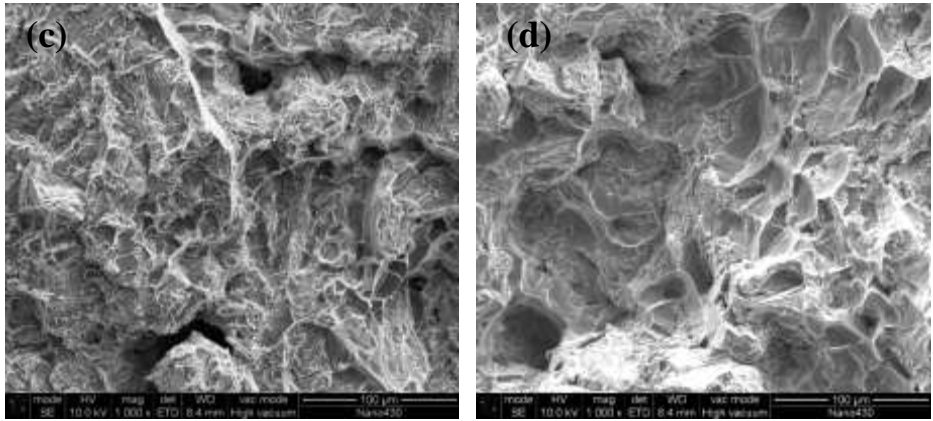

**Fig.5.** SEM images of the fracture surface of tensile test alloys: (a) Non-treated; (b) P; (c) UT; (d) P+UT.

Table 1 mechanical property for the four applied methods.

| Method | Ultimate tensile stress (MPa) | Yield stress (MPa) | Elongation (%) |
|---|---|---|---|
| Non-treated | 163±10 | 121±2 | 4.2±0.8 |
| P | 248±7 | 182±5 | 12.8±0.8 |
| UT | 232±4 | 169±9 | 5.3±0.7 |
| P+UT | 268±9 | 192±1 | 17.1±1.6 |

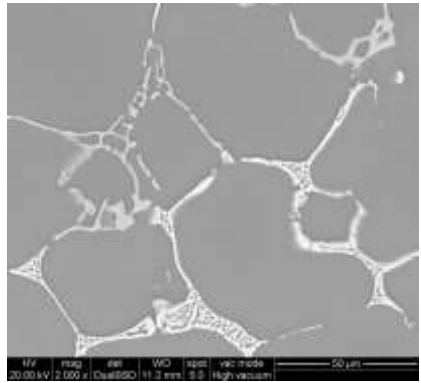